\documentclass[reprint,
    aps,prl,
    superscriptaddress,
    longbibliography,
    amsmath,amssymb,
    balancelastpage
]{revtex4-2}

\usepackage[T1]{fontenc}
\usepackage[utf8]{inputenc}
\usepackage{graphicx}
\usepackage{siunitx}
\usepackage{mathptmx}
\usepackage{bm}
\usepackage{braket}
\usepackage{ORCIDinREVTeX}
\usepackage{calrsfs}
\DeclareMathAlphabet{\pazocal}{OMS}{zplm}{m}{n}

\usepackage[unicode=true,
    colorlinks=true,
    linkcolor=blue,         % color of internal links (change box color with linkbordercolor)
    citecolor=blue,         % color of links to bibliography
    filecolor=blue,         % color of file links
    urlcolor=blue]{hyperref}
    
\renewcommand{\i}{\text{i}}
\renewcommand{\eqref}[1]{\hyperref[{#1}]{\textup{(\ref*{#1}})}}
\newcommand{\secref}[1]{\hyperref[{#1}]{\textup{Sec.~\ref*{#1}}}}
\newcommand{\tabref}[1]{\hyperref[{#1}]{\textup{Table~\ref*{#1}}}}
\newcommand{\appref}[1]{\hyperref[{#1}]{\textup{Appendix~\ref*{#1}}}}

\newcommand{\vect}[1]{\mathbf{#1}}          % vector
\newcommand{\highlight}[1]{\textcolor{blue}{#1}}

\makeatletter
\newcommand\figref[1]{%
    \@ifnextchar\bgroup{\figref@double{#1}}{\figref@single{#1}}%
}
\newcommand\figref@single[1]{\hyperref[{#1}]{\textup{FIG.~\ref*{#1}}}~}
\newcommand\figref@double[2]{\hyperref[{#1}]{\textup{FIG.~\ref*{#1}}\textcolor{blue}{\textbf{#2}}}}
\makeatother

\allowdisplaybreaks
\begin{document}

\title{Topological invariant of non-Hermitian space-time modulated photonic crystals}

\author{Xiaoke Gao}
\orcid{0000-0002-3769-4451}

\author{Xiaoyu Zhao}
\orcid{0000-0001-8901-7498}

\author{Jiawei Wang}
\orcid{0000-0003-1526-110X}

\author{Xikui Ma}
\orcid{0000-0003-4138-7443}

\author{Tianyu Dong}
\orcid{0000-0003-4816-0073}
% \homepage{http://www.donglab.cn/en/}
\email[Corresponding author. Email: ]{tydong@mail.xjtu.edu.cn}
\affiliation{School of Electrical Engineering, Xi'an Jiaotong University, Xi'an 710049, China.}%

\date{December 29, 2024}

% - Abstract
\begin{abstract}
We propose a medium transformation approach to formulate the adjoint system of space-time modulated photonic crystals (STMPCs), essential for the bi-orthogonal Berry connection when calculating the topological invariant. We show that the non-Abelian Zak phase of STMPCs comprising stacked photonic time crystals and dielectrics is quantized to 0 or 1 for both the entangled and isolated bands. We find that the eigenmodes at the center and edge of the Brillouin zone differ in symmetry for the band with non-trivial Zak phases, while they share the same symmetry for the trivial Zak phases. In addition, topological phase transitions owing to band inversion are observed. Moreover, a generalized Brillouin zone of the non-Hermitian STMPCs is established, which is identical to the Hermitian counterpart, implicating that the non-Bloch band theory is not required in this regard. The proposed medium transformation method may serve as an alternative approach to exploring more intricate topological phenomena in non-Hermitian systems when incorporating non-Bloch band theory.
\end{abstract}

\keywords{topological invariant, photonic crystals, space-time modulation, Zak phase, band structure}

\maketitle   % Comment out to count words --

% \paragraph{Introduction}
The exploration of control of wave behavior through the spatial and temporal dimension has yielded fascinating conceptual applications \emph{e.g.}, energy manipulation \cite{lee2021parametric,yang2023cascaded}, frequency shift \cite{wang2023broadband,moreno2024space,zhao2024controllable}, thermal design \cite{torrent2018nonreciprocal} and beam steering \cite{biancalana2007dynamics,cong2020spatiotemporal} \emph{etc}. Space-time modulated photonic crystals (STMPCs) provide an exceptional optical platform \cite{sharabi2022spatiotemporal,park2021spatiotemporal}, which may be categorized into two types based on the interaction of temporal and spatial modulation and the discrete sets of eigenmodes: One is spatiotemporal photonic crystals (PCs) \cite{ding2024non} with a traveling wavelike modulation, characterized by material parameters ($\varepsilon, \mu$) with the perturbation $\delta\varepsilon$ proportional to $\sin(g x - \Omega t + \phi_0)$, where $g$ and $\Omega$ respectively denote the spatial and temporal reciprocal lattice, and $\phi_0$ denotes the initial phase of modulation; the other type of PCs has material parameters ($\varepsilon, \mu$) in a separable form of $\varepsilon(x, t) = f(x) h(t)$ where $f(x)$ and $h(t)$ are periodic functions \cite{sharabi2022spatiotemporal}, which is often referred to as space-time periodic PCs. For spatiotemporal PCs, frequency $\omega$ and momentum $k$ constitute a valid quantum number $(\omega, k)$, rather than being considered separately \cite{galiffi2022photonics}, indicating that each eigenmode is characterized by frequency-momentum vectors $(k + n g, \omega + n \Omega)$, where $n \in \mathbb{Z}$. For space-time periodic PCs, the decoupling of temporal and spatial modulations results in two distinct quantum numbers, allowing the characterization of eigenmodes by frequency-momentum vectors $(k + n g, \omega + m \Omega)$, where $n \in \mathbb{Z}$ and $m \in \mathbb{Z}$. The fundamental solutions of these two types of STMPCs may be expressed in terms of Floquet-Bloch modes owing to periodic modulations in spatial and temporal dimensions. This allows for the analysis of properties such as non-reciprocity \cite{sounas2017non}, exceptional points \cite{kazemi2019exceptional}, spectral degeneracies \cite{park2021spatiotemporal}, topological features \cite{lustig2018topological,lu2021floquet}, \emph{etc.}, within the band theory.

The topological properties of PCs are fundamentally important. When temporal modulation is introduced to PCs, complex bands with exceptional points and degeneracy appear because of non-Hermiticity, which challenges the investigation of topological properties. In a simple scenario concerning only temporal modulation, e.g. for photonic time crystals (PTCs), the topological properties can be obtained from traditional PCs through space-time duality \cite{xiao2014surface,lustig2018topological}, thereby bypassing the double $k$-degeneracy of the momentum band gap. In parity-time ($\pazocal{PT}$)-symmetric PCs, the conventional non-Abelian Zak phase is quantized only in the $\pazocal{PT}$-exact phase, whereas the topological characteristics in the $\pazocal{PT}$-broken phase cannot be determined using the same method \cite{ding2015coalescence}. For an STMPC consisting of PTCs and dielectrics, the Zak phase is calculated for the isolated bands, whereas the situation involving entangled bands resulting from the momentum band gap is omitted \cite{lu2021floquet}. The advancement of non-Bloch band theory has allowed the investigation of the topological characteristics of the aforementioned non-Hermitian crystals \cite{yao2018edge,song2019non,yokomizo2019non}. By introducing the generalized Brillouin zone, topological invariants are redefined \cite{yokomizo2022non,yokomizo2024non}, and a new bulk-edge correspondence principle is established that can be used to explain the non-Hermitian skin effect. Recent research has indicated that non-Bloch band theory can be applied effectively to chiral metamaterials \cite{yokomizo2024non} and PCs with periodically aligned gain and loss \cite{yokomizo2022non}. In addition, by virtue of effective medium theory \cite{huidobro2021homogenization}, a thorough non-Bloch band theory can be formulated for spatiotemporal PCs only in the long-wavelength limit \cite{ding2024non}, indicating that further efforts are required to investigate the topological characteristics of STMPCs.

In this Letter, we present a medium transformation approach to formulate the adjoint system for STMPCs, which is essential for the bi-orthogonal Berry connection when calculating topological invariants and is applicable throughout the whole spectrum. We reveal that the Zak phase of an STMPC consists of regularly stacked PTCs and dielectrics, is quantized to either zero or one for both the entangled and isolated bands, and this has been validated by examining the correlation between the Zak phase and the symmetry of the eigenmodes and by observing the topological phase transition resulting from band inversion. In addition, a generalized Brillouin zone for STMPCs has been established, which is identical to that of Hermitian systems. This indicates that the non-Bloch theory is not strictly necessary for space-time decoupled modulated PCs without material loss and external gain, and traditional Bloch band theory may still be applicable.

\paragraph{Material Transformation} 
As non-Hermitian systems, STMPCs exhibit complex band structures, with the eigenstates corresponding to each band not being orthogonal. Now, the bilinear product should be used to calculate the biorthogonal Berry connection for the band $n$, \emph{i.e.}, $\pazocal{A}_n(\vect{k}) = \text{i} \Braket{\psi^\text{L}_{n,\vect{k}} | \nabla_{\vect{k}} | \psi^{\text{R}}_{n,\vect{k}}}$, where $\psi^{\text{L/R}}_{n,\vect{k}}$ denotes the left or right eigenstates associated with the point $\vect{k}$ on the band $n$. Often, symmetries of the system can be utilized to obtain the relationship between the left and right eigenstates, offering an indirect approach for PCs. In particular, for PCs stacked one-dimensional, the transfer matrix method (TMM) can be used to calculate the right eigenstates $\psi^{\text{R}}_{n,\vect{k}}$, while obtaining the left eigenstates is not straightforward for STMPCs. Here, we provide a medium transformation method to calculate the left eigenstates using the TMM. 

\begin{figure}[!htb]    % only used for initial submission
    \centering
    \includegraphics[width = 0.4\textwidth]{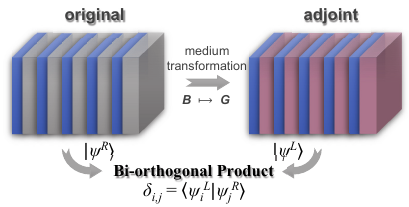}
    \caption{Theoretical framework diagram of material transformation.}
    \label{fig:figure01}
\end{figure}

Considering the periodic modulation in both the space and time domain, Maxwell equations can be written as
\begin{equation} \label{eq:maxwellBF}
    \pazocal{L}\psi^{\text{R}} - \text{i} \partial_t (\bm{B} \psi^{\text{R}}) = \omega \bm{B} \psi^{\text{R}}
\end{equation}
under the Floquet--Bloch ansatz with quasi-frequency $\omega$, where the operator $\pazocal{L}$ and material matrix respectively reads
\begin{align}
    \pazocal{L} &= e^{-\text{i} \vect{k} \cdot \vect{r}} 
    \begin{pmatrix}
        0 & \text{i} \nabla \times \\
        -\text{i} \nabla \times & 0
    \end{pmatrix}
    e^{\text{i} \vect{k} \cdot \vect{r}}, \\
    \bm{B} &= 
    \begin{pmatrix}
        \bm{\varepsilon} & \bm{\xi} \\
        \bm{\zeta}       & \bm{\mu} 
    \end{pmatrix}.
\end{align}
Here, $\psi^{\text{R}} = (\bm{e}, \bm{h})^{\text{T}}$ denotes the amplitudes of the electric field $\bm{E}(\vect{r},t) = \bm{e}(\vect{r},t) e^{\text{i} (\vect{k}\cdot\vect{r} - \omega t)}$ and the magnetic field $\bm{H}(\vect{r},t) = \bm{h}(\vect{r},t) e^{\text{i} (\vect{k}\cdot\vect{r} - \omega t)}$ with $\vect{k}$ being the quasi-momentum. Since the operators $\pazocal{L}$ and $\text{i} \partial_t$ both are Hermitian, the adjoint system equation of \eqref{eq:maxwellBF} can be expressed as $\pazocal{L}\psi^\text{L} - \bm{B}^\dagger (\text{i} \partial_t \psi^\text{L}) = \omega^* \bm{B}^\dagger \psi^\text{L}$, which can be reformulated as
\begin{equation} \label{eq:maxwellBFadjG}
    \pazocal{L}\psi^\text{L} - \text{i} \partial_t ( \bm{G}\psi^\text{L}) = \omega^* \bm{G} \psi^\text{L},
\end{equation}
with the transformed material $\bm{G}$ satisfying
\begin{equation} \label{eq:Operatortransform}
    (\text{i} \partial_t + \omega^*)(\bm{G}\psi^\text{L}) = \bm{B}^\dagger (\text{i} \partial_t + \omega^*)\psi^\text{L}.
\end{equation}
Here, the asterisk denotes the conjugate and the dagger denotes the conjugate transpose. Therefore, we can conclude that the left eigenstates $\psi^\text{L}$ of the adjoint system \eqref{eq:maxwellBFadjG} are the right eigenstates of the original system \eqref{eq:maxwellBF} with the transformed material $\bm{G}$, the same quasi-momentum $\vect{k}$, and the conjugate quasi-frequency $\omega^*$, as illustrated in \figref{fig:figure01}{}. Consequently, by applying the TMM to the transformed material $\bm{G}$, the solution to the adjoint system $\eqref{eq:maxwellBFadjG}$ can be derived.

Next, we present a strategy to derive the transformed material $\bm{G}$. Using the Fourier series expansion, we can obtain $\bm{B}^\dagger = \sum_{p}\bm{B}^\dagger_p e^{-\text{i} p \Omega t}$ and $\psi^\text{L} = \sum_{m} \ket{\psi_m^\text{L}}= \sum_{m}\psi_m^\text{L} e^{-\text{i} m \Omega t}$, where $\Omega$ denotes the fundamental frequency. Likewise, the adjoint material $\bm{G}$ may also be expressed in terms of series as
\begin{equation} \label{eq:adjMaterialExpansion}
    \bm{G} = \sum_{q = -\infty}^{\infty} \sum_{m = -\infty}^{\infty}\bm{G}_{q m} e^{-\text{i} (q-m) \Omega t}.
\end{equation}
where $\bm{G}_{qm}$ corresponds to the conversion of the $m$-th harmonic mode $\ket{\psi^{\text{L}}_m}$ to the $q$-th harmonic mode $\ket{\psi^{\text{L}}_q}$. Note that the spatial variable $\vect{r}$ is ignored hereafter for convenience. Now, by virtue of the Cauchy product of series, the equation of material transformation \eqref{eq:Operatortransform} can be reformulated as \cite{supplementary}
\begin{equation} \label{eq:Seriestransform2}
    \sum_{m}(q\Omega+\omega^*)\bm{G}_{q m}\psi^\text{L}_m = 
    \sum_{n}(n\Omega+\omega^*) \bm{B}^\dagger_{q-n} \psi^\text{L}_n,
\end{equation}
for each $q \in \mathbb{Z}$.
Often a limited number of terms in the Fourier series expansion is necessary \cite{lu2021floquet,park2021spatiotemporal}, \emph{i.e.}, the integers $m,n,q \in [-N, N]$ in \eqref{eq:Seriestransform2}. As a result, we can derive a sufficient solution for the adjoint material, which reads
\begin{equation} \label{eq:Elementtransform}
    \bm{G}_{qn} = (q\Omega+\omega^*)^{-1} \bm{B}^\dagger_{q-n} (n\Omega+\omega^*)
\end{equation} 
Now, $\bm{G}$ can be reconstructed according to \eqref{eq:adjMaterialExpansion}; consequently, the eigen states of the adjoint system can be derived.

\begin{figure}[!htb]    % only used for initial submission
    \centering
    \includegraphics[width = 0.45\textwidth]{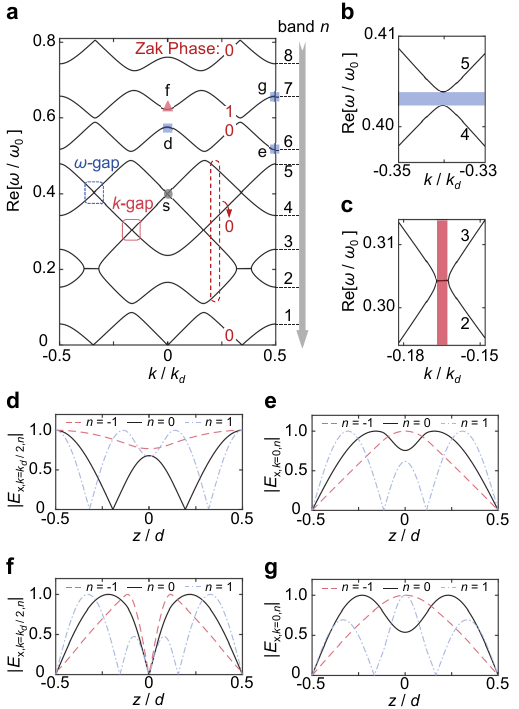}
    \caption{\textbf{Topological characteristics of the complex band structure of the STMPCs.} (\textbf{a}) Real component of the band structure. The Zak phases are denoted by integers; `s' represents the twice-degenerated point; the triangle markers `d' and cubic markers `e-g' signify the eigenstates with distinct symmetries at the center or edge of the band, corresponding to the distribution of the normalized electric field $|E_{x,k,n}|$ in (\textbf{d}) -- (\textbf{g}), respectively. The band number is indicated to the right of the figure. (\textbf{b}) and (\textbf{c}) depict the zoomed views of the $\omega$-gap and $k$-gap, indicated by dashed and solid boxes in (\textbf{a}), respectively. The blue and red shaded areas denote the energy gap and momentum gap, respectively. (\textbf{d}--\textbf{g}) The distribution of the normalized electric field $|E_{x,k,n}|$ corresponding to the eigenstates labeled `d-g' in (\textbf{a}). The dashed, solid, and dotted lines indicate the harmonics with the order of $n = -1, 0, 1$, respectively. In our calculation, the filling ratio and modulation frequency of the STMPC read $\eta = 0.25$ and $\Omega = 0.4 \omega_0$, respectively; all other parameters are the same to those in \highlight{Figure S5} in \highlight{Supplemental Material, S4}.}
    \label{fig:figure02}
\end{figure}

\paragraph{Zak Phase of an STMPC}
Time modulation introduces non-Hermitian characteristics, resulting in entanglement and degeneracy of the bands, which requires the bi-orthogonal bases $\psi^\text{L}_{i,k}$ and $\psi^\text{R}_{i,k}$, and the overlap matrices $\pazocal{U}(k)$ (see in \highlight{Supplemental Material, S3} for the explicit forms of $\pazocal{U}(k)$) to calculate the non-Abelian Zak phase. \figref{fig:figure02} illustrates the non-Abelian Zak phase of an STMPC. Here, bands 6, 7, and 8 are isolated, with their Zak phases indicated in red, 0 or 1, as shown in \figref{fig:figure02}{(a)}. Similarly to traditional PTCs, bands 2 and 3 are entangled since the $k$-gap at half the modulation frequency $\omega = \Omega / 2$; however, the entanglement between bands 3 and 4 is due to the $k$-gap near $\omega = 0.3 \omega_0$, which is not observed in traditional PTCs; moreover, bands 4 and 5 entangle when $k=0$ and $\omega = \Omega$ (denoted by the point `s', see the gray dot in \figref{fig:figure02}{(a)}). For the entangled bands, individual Zak phase cannot be clearly defined due to band crossings or degeneracy, while a total Zak phase for the entire group of entangled bands can be calculated according to \cite{ding2015coalescence,supplementary,blanco2020tutorial}
 \begin{equation}
     \theta = -\Im \left[ \int_{\partial B}  \ln \left( \text{det} \Braket{\psi^\text{L}_{i,k} | \partial_k \psi^\text{R}_{j,k}} \right) \mathrm{d}k \right],
 \end{equation}
where $\Im[\bullet]$ denotes the imaginary part; the integral interval $\partial B$ denotes the first Brillouin region and the subscripts $i$ and $j$ represent the number of the entangled bands, \emph{i.e.}, $i,j \in \{2,3,4,5\}$, which have a total Zak phase of $\theta = 0$, indicated by the number in \figref{fig:figure02}{(a)}. However, in a typical non-Hermitian system with PT symmetry, previous research has revealed that in the PT-broken phase, the non-Abelian Zak phase for the entangled bands fails to be a topological invariant \cite{ding2015coalescence,stegmaier2021topological}, demanding the non-Bloch band theorem \cite{ding2024non,yokomizo2024non}. The calculation result presented here demonstrates that the previous definition and numerical method of the non-Abelian Zak phase \cite{yang2024non} remain applicable to STMPCs; although both PT-symmetric and space-time-varying systems exhibit band entanglement, they differ from each other from a topological perspective. \figref{fig:figure02}{(b,c)} show the zoom-in view of the $\omega$-gap and $k$-gap highlighted by the dashed and solid boxes in \figref{fig:figure02}{(a)}, respectively, which correspond to the energy gap resulting from spatial modulation and the momentum gap due to temporal modulation. \figref{fig:figure02}{(d--g)} depict the distribution of the electric field $|E_{x,k,n}|$ associated with eigenstates at the center or edge of the band structure. Unlike conventional PCs, each STMPC eigenstate has several time harmonics, which have been truncated to $n=-1,0,1$ for sinusoidal modulation in the temporal dimension (in \highlight{Supplemental Material, S1})\cite{supplementary}. The different symmetries correspond to conditions $|E_{x,k,n}(z=0)| \neq 0$ and $|E_{x,k,n}(z=0)| = 0$, which are also indicated by the cubes and the triangle in \figref{fig:figure02}{(a)}, respectively. For the isolated band 6 with Zak phase $\theta_6 = 0$, the field distributions when $k = 0$ and $k = k_d / 2$ exhibit identical symmetry, as seen in \figref{fig:figure02}{(d,e)}. However, for the isolated band $7$ with Zak phase $\theta_7 = 1$, the field distributions when $k = 0$ and $k = k_d / 2$ exhibit different symmetry, as seen in \figref{fig:figure02}{(f,g)}. The relationship between the Zak phase and the symmetry of isolated bands in STMPCs is consistent with the findings of previous studies on spatially periodic systems exhibiting inversion symmetry, such as periodic solids \cite{zak1989berry}, PCs \cite{xiao2014surface}, and periodic acoustic systems \cite{xiao2015geometric}.  

\begin{figure}[!htb]    % only used for initial submission
    \centering
    \includegraphics[width = 0.45\textwidth]{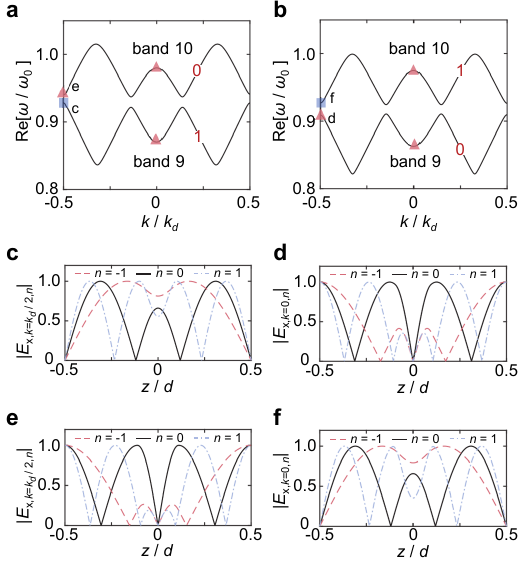}
    \caption{\textbf{Band inversion within isolated bands.} Band structures of (\textbf{a}) $9$-th and (\textbf{b}) $10$-th bands when the filling ratio are $\eta_1 = 0.19$ and $\eta_2 = 0.22$, respectively. Numbers on the bands labels the Zak phases. The triangle and cubic markers signify the eigenstates with distinct symmetries at the band center or edge. (\textbf{c}--\textbf{f}) The distribution of the normalized electric fields $|E_{x,k,n}|$ corresponds to the eigenstates labeled `c--f', respectively, in (\textbf{a}) and (\textbf{b}). The dashed, solid, and dotted lines indicate the harmonics with the order $n = -1, 0, 1$, respectively. In the calculation, the modulation frequency reads $\Omega = 0.4\omega_0$.}
    \label{fig:figure03}
\end{figure}

\paragraph{Band Inversion Phenomenon}
Upon the correlation between the Zak phase and the symmetry characteristics of the edge states is established, the modulation of the Zak phase can be facilitated using the band inversion. \figref{fig:figure03}{(a)} and \figref{fig:figure03}{(b)} depict isolated bands $9$ and $10$ of STMPCs with filling ratios $\eta_1 = 0.19$ and $\eta_2 = 0.22$, respectively, showing the phenomenon of band inversion. Although the band structure changes little when the filling ratio changes, the symmetry characteristics of the edge modes change, whether $|E_{x,k,n}(z=0)| = 0$ or not, as indicated by triangles and cubes. By modifying the filling ratio $\eta_1$ to $\eta_2$, the Zak phase transition can be observed, \emph{e.g.}, $\theta_9$ changing from 1 to 0 for band 9 (or $\theta_{10}$ from $0$ to $1$ for band $10$). \figref{fig:figure03}{(c--f)} shows the field distribution corresponding, respectively, to the points labeled `c--f'  in \figref{fig:figure03}{(a)} and \figref{fig:figure03}{(b)}. The field distributions of `c' (or `e') closely resemble those of `f' (or `d'), which indicate an exchange of eigenstates for bands $9$ and $10$ at the band edge, hence resulting in a symmetry inversion at the band edge. The eigenstates near the band center are barely influenced by the small changes in the filling ratio; in addition, the substantial energy gap inhibits the interchange of eigenstates, hence preserving the symmetry characteristics. Consequently, for STMPCs that still retain spatial inversion symmetry, the band inversion phenomenon \cite{zak1989berry,xiao2014surface} can be directly generalized to explain the topological transition. Moreover, the phenomenon of band inversion also occurs between isolated bands and degenerate bands (see in \highlight{Supplemental Material, S5}), implying that the transition of the Zak phase of the STMPCs may be explained by the band inversion proposed by Zak \cite{zak1989berry,supplementary}, regardless of whether the eigenstate is located within isolated or entangled bands.

\begin{figure}[!htb]    % only used for initial submission
    \centering
    \includegraphics[width = 0.45\textwidth]{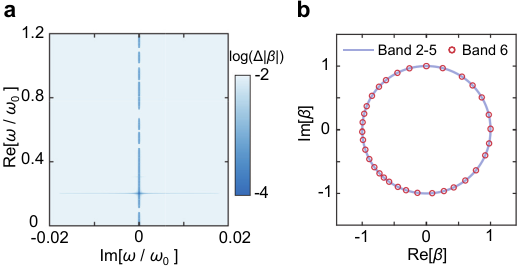}
    \caption{\textbf{Spectrum under non-Bloch band theory framework.} (\textbf{a}) Continuum spectra of the STMPCs under open boundary conditions in the thermodynamic limit. (\textbf{b}) Generalized Brillouin zones of the entangled bands 2--5 (line) and the isolated band 6 (circles). In the calculation, the parameters are the same as those of \hyperref[fig:figure02]{FIG. 2}.}
    \label{fig:figure04}
\end{figure}

\paragraph{Validation within Non-Bloch Band Theory}
Non-Bloch band theory has been widely employed to elucidate the topological characteristics of non-Hermitian systems with intriguing band structures that may exhibit exceptional degeneracy. Here, we demonstrate that the topological characteristics can be effectively described using both traditional Bloch and non-Bloch band theories for non-Hermitian systems. Within the non-Bloch band theory, the generalized Brillouin zone can be determined by solving the eigenvalue problem $f(\beta, \omega) = 0$ resulting from the transfer matrix for a single unit cell ($\beta = e^{\text{i} k d}$) \cite{supplementary}, which is subject to the constraint $|\beta_{2N_t+1}| = |\beta_{2N_t+2}|$, where $\beta_{2N_t+1}$ and $\beta_{2N_t+2}$ represent the $(2N_t+1)$-th and $(2N_t+2)$-th eigenvalues, respectively\cite{yokomizo2022non,yokomizo2024non}. Note that eigenvalues are generally arranged by their moduli. \figref{fig:figure04}{(a)} depicts the eigenspectra (dark blue line) of the STMPC, for which $\Delta|\beta| = \left||\beta_{2N_t+1}| - |\beta_{2N_t+2}| \right|$ is used. The eigenspectra $\omega$ exhibit complex frequency right in the momentum gaps located at half the modulated frequency, that is, when $\Re[\omega/\omega_0] = 0.2$ and $\Re[\omega/\omega_0] = 0.304$, as indicated by the solid box in \figref{fig:figure02}{(a)}. \figref{fig:figure04}{(b)} displays the generalized Brillouin zone, which represents the trajectory of the eigenvalue $\beta_{2N_t+1}$ associated with the eigenspectra $\omega$ in \figref{fig:figure04}{(a)}. The radii of generalized Brillouin zone for both the entangled bands (line) and isolated bands (circles) are equal to one, which are identical to the normal Brillouin zone used in Hermitian systems. Consequently, the integral region used to calculate the Zak phase within the non-Bloch theory framework \cite{ding2024non} is the same as that used in the Hermitian context, indicating that the Zak phase of such STMPCs can also be determined using the conventional Bloch theory. In particular, for the STMPCs considered constituting a periodic stack of pure PTCs and conventional dielectrics, the decoupling of temporal and spatial modulations results in spectra that are entirely real or consist of conjugate pairs, distinguishing it from the wave-like STMPCs \cite{ding2024non,sharabi2022spatiotemporal}. Furthermore, the system considered achieves non-Hermiticity by exploiting the non-commutativity between $\partial_t$ and $\varepsilon(t)$, without incorporating medium gain or loss, but exclusively by adjusting the real component of the permittivity, which results in the generalized Brillouin zone of such a non-Hermitian system being identical to that of a Hermitian system. Therefore, when analyzing the topological characteristics, non-Bloch band theory is not necessarily required, and the conventional Bloch band theory may also be effective.

% \paragraph{Conclusion}
In summary, we have proposed a material transformation strategy to construct the adjoint system of a space-time-modulated photonic crystal (STMPC). The eigenstates of the adjoint system and the original system provide a complete biorthogonal basis, enabling a straightforward resolution of the non-Abelian Zak phase without the necessity of effective approximation methods. In the case of space-time decoupled modulation, the generalized Brillouin zone of a non-Hermitian system is identical to that of the Hermitian PC, namely $|\beta| = 1$, indicating that the conventional Bloch theory framework may still be applicable. Furthermore, the relationship between the Zak phase and the symmetry characteristics of the edge states remains valid, and the topological transition induced by band inversion can still be observed. Moreover, by combining with non-Bloch band theory, the proposed material transformation method for space-time-decoupled modulated PCs can potentially serve as an alternative method for investigating more complex non-Hermitian systems.

\paragraph*{Acknowledgments}
T.D. acknowledges the support of the National Natural Science Foundation of China (NSFC) under grant no. 51977165. X.G. thanks the support from the Fundamental Research Funds for the Central Universities (xzy022023035).

\paragraph*{Data and Code Availability}
All data needed to evaluate the conclusions are presented in the article and/or the Supplemental Material \cite{supplementary}. Raw data and the relevant code are available from the corresponding author on request.

% \paragraph*{Competing Interests}
% The authors declare that they have no competing interests. 

% \renewcommand*{\bibfont}{\small}
% \setlength{\bibsep}{0.5ex plus 0.5ex}
\bibliography{Main}

\begin{thebibliography}{30}%
\makeatletter
\providecommand \@ifxundefined [1]{%
 \@ifx{#1\undefined}
}%
\providecommand \@ifnum [1]{%
 \ifnum #1\expandafter \@firstoftwo
 \else \expandafter \@secondoftwo
 \fi
}%
\providecommand \@ifx [1]{%
 \ifx #1\expandafter \@firstoftwo
 \else \expandafter \@secondoftwo
 \fi
}%
\providecommand \natexlab [1]{#1}%
\providecommand \enquote  [1]{``#1''}%
\providecommand \bibnamefont  [1]{#1}%
\providecommand \bibfnamefont [1]{#1}%
\providecommand \citenamefont [1]{#1}%
\providecommand \href@noop [0]{\@secondoftwo}%
\providecommand \href [0]{\begingroup \@sanitize@url \@href}%
\providecommand \@href[1]{\@@startlink{#1}\@@href}%
\providecommand \@@href[1]{\endgroup#1\@@endlink}%
\providecommand \@sanitize@url [0]{\catcode `\\12\catcode `\$12\catcode `\&12\catcode `\#12\catcode `\^12\catcode `\_12\catcode `\%12\relax}%
\providecommand \@@startlink[1]{}%
\providecommand \@@endlink[0]{}%
\providecommand \url  [0]{\begingroup\@sanitize@url \@url }%
\providecommand \@url [1]{\endgroup\@href {#1}{\urlprefix }}%
\providecommand \urlprefix  [0]{URL }%
\providecommand \Eprint [0]{\href }%
\providecommand \doibase [0]{https://doi.org/}%
\providecommand \selectlanguage [0]{\@gobble}%
\providecommand \bibinfo  [0]{\@secondoftwo}%
\providecommand \bibfield  [0]{\@secondoftwo}%
\providecommand \translation [1]{[#1]}%
\providecommand \BibitemOpen [0]{}%
\providecommand \bibitemStop [0]{}%
\providecommand \bibitemNoStop [0]{.\EOS\space}%
\providecommand \EOS [0]{\spacefactor3000\relax}%
\providecommand \BibitemShut  [1]{\csname bibitem#1\endcsname}%
\let\auto@bib@innerbib\@empty
%</preamble>
\bibitem [{\citenamefont {Lee}\ \emph {et~al.}(2021)\citenamefont {Lee}, \citenamefont {Park}, \citenamefont {Cho}, \citenamefont {Wang}, \citenamefont {Kim}, \citenamefont {Daraio},\ and\ \citenamefont {Min}}]{lee2021parametric}%
  \BibitemOpen
  \bibfield  {author} {\bibinfo {author} {\bibfnamefont {S.}~\bibnamefont {Lee}}, \bibinfo {author} {\bibfnamefont {J.}~\bibnamefont {Park}}, \bibinfo {author} {\bibfnamefont {H.}~\bibnamefont {Cho}}, \bibinfo {author} {\bibfnamefont {Y.}~\bibnamefont {Wang}}, \bibinfo {author} {\bibfnamefont {B.}~\bibnamefont {Kim}}, \bibinfo {author} {\bibfnamefont {C.}~\bibnamefont {Daraio}},\ and\ \bibinfo {author} {\bibfnamefont {B.}~\bibnamefont {Min}},\ }\bibfield  {title} {\bibinfo {title} {Parametric oscillation of electromagnetic waves in momentum band gaps of a spatiotemporal crystal},\ }\href {https://doi.org/10.1364/PRJ.406215} {\bibfield  {journal} {\bibinfo  {journal} {Photonics Res.}\ }\textbf {\bibinfo {volume} {9}},\ \bibinfo {pages} {142} (\bibinfo {year} {2021})}\BibitemShut {NoStop}%
\bibitem [{\citenamefont {Yang}\ \emph {et~al.}(2023)\citenamefont {Yang}, \citenamefont {Hu}, \citenamefont {Li},\ and\ \citenamefont {Luo}}]{yang2023cascaded}%
  \BibitemOpen
  \bibfield  {author} {\bibinfo {author} {\bibfnamefont {Q.}~\bibnamefont {Yang}}, \bibinfo {author} {\bibfnamefont {H.}~\bibnamefont {Hu}}, \bibinfo {author} {\bibfnamefont {X.}~\bibnamefont {Li}},\ and\ \bibinfo {author} {\bibfnamefont {Y.}~\bibnamefont {Luo}},\ }\bibfield  {title} {\bibinfo {title} {Cascaded parametric amplification based on spatiotemporal modulations},\ }\href {https://doi.org/10.1364/PRJ.472233} {\bibfield  {journal} {\bibinfo  {journal} {Photonics Res.}\ }\textbf {\bibinfo {volume} {11}},\ \bibinfo {pages} {B125} (\bibinfo {year} {2023})}\BibitemShut {NoStop}%
\bibitem [{\citenamefont {Wang}\ and\ \citenamefont {Wang}(2023)}]{wang2023broadband}%
  \BibitemOpen
  \bibfield  {author} {\bibinfo {author} {\bibfnamefont {N.}~\bibnamefont {Wang}}\ and\ \bibinfo {author} {\bibfnamefont {G.~P.}\ \bibnamefont {Wang}},\ }\bibfield  {title} {\bibinfo {title} {Broadband frequency translation by space--time interface with weak permittivity temporal change},\ }\href {https://doi.org/10.1364/OL.494957} {\bibfield  {journal} {\bibinfo  {journal} {Opt. Lett.}\ }\textbf {\bibinfo {volume} {48}},\ \bibinfo {pages} {4436} (\bibinfo {year} {2023})}\BibitemShut {NoStop}%
\bibitem [{\citenamefont {Moreno-Rodr\'{\i}guez}\ \emph {et~al.}(2024)\citenamefont {Moreno-Rodr\'{\i}guez}, \citenamefont {Alex-Amor}, \citenamefont {Padilla}, \citenamefont {Valenzuela-Vald\'es},\ and\ \citenamefont {Molero}}]{moreno2024space}%
  \BibitemOpen
  \bibfield  {author} {\bibinfo {author} {\bibfnamefont {S.}~\bibnamefont {Moreno-Rodr\'{\i}guez}}, \bibinfo {author} {\bibfnamefont {A.}~\bibnamefont {Alex-Amor}}, \bibinfo {author} {\bibfnamefont {P.}~\bibnamefont {Padilla}}, \bibinfo {author} {\bibfnamefont {J.~F.}\ \bibnamefont {Valenzuela-Vald\'es}},\ and\ \bibinfo {author} {\bibfnamefont {C.}~\bibnamefont {Molero}},\ }\bibfield  {title} {\bibinfo {title} {Space-time metallic metasurfaces for frequency conversion and beamforming},\ }\href {https://doi.org/10.1103/PhysRevApplied.21.064018} {\bibfield  {journal} {\bibinfo  {journal} {Phys. Rev. Appl.}\ }\textbf {\bibinfo {volume} {21}},\ \bibinfo {pages} {064018} (\bibinfo {year} {2024})}\BibitemShut {NoStop}%
\bibitem [{\citenamefont {Zhao}\ \emph {et~al.}(2024)\citenamefont {Zhao}, \citenamefont {Gao}, \citenamefont {Wang}, \citenamefont {Ma},\ and\ \citenamefont {Dong}}]{zhao2024controllable}%
  \BibitemOpen
  \bibfield  {author} {\bibinfo {author} {\bibfnamefont {X.}~\bibnamefont {Zhao}}, \bibinfo {author} {\bibfnamefont {X.}~\bibnamefont {Gao}}, \bibinfo {author} {\bibfnamefont {J.}~\bibnamefont {Wang}}, \bibinfo {author} {\bibfnamefont {X.}~\bibnamefont {Ma}},\ and\ \bibinfo {author} {\bibfnamefont {T.}~\bibnamefont {Dong}},\ }\bibfield  {title} {\bibinfo {title} {Controllable location-dependent frequency conversion based on space-time transformation optics},\ }\href {https://doi.org/10.1088/1361-6463/ad6d7b} {\bibfield  {journal} {\bibinfo  {journal} {J. Phys. D: Appl. Phys.}\ }\textbf {\bibinfo {volume} {57}},\ \bibinfo {pages} {455103} (\bibinfo {year} {2024})}\BibitemShut {NoStop}%
\bibitem [{\citenamefont {Torrent}\ \emph {et~al.}(2018)\citenamefont {Torrent}, \citenamefont {Poncelet},\ and\ \citenamefont {Batsale}}]{torrent2018nonreciprocal}%
  \BibitemOpen
  \bibfield  {author} {\bibinfo {author} {\bibfnamefont {D.}~\bibnamefont {Torrent}}, \bibinfo {author} {\bibfnamefont {O.}~\bibnamefont {Poncelet}},\ and\ \bibinfo {author} {\bibfnamefont {J.-C.}\ \bibnamefont {Batsale}},\ }\bibfield  {title} {\bibinfo {title} {Nonreciprocal thermal material by spatiotemporal modulation},\ }\href {https://doi.org/10.1103/PhysRevLett.120.125501} {\bibfield  {journal} {\bibinfo  {journal} {Phys. Rev. Lett.}\ }\textbf {\bibinfo {volume} {120}},\ \bibinfo {pages} {125501} (\bibinfo {year} {2018})}\BibitemShut {NoStop}%
\bibitem [{\citenamefont {Biancalana}\ \emph {et~al.}(2007)\citenamefont {Biancalana}, \citenamefont {Amann}, \citenamefont {Uskov},\ and\ \citenamefont {O'Reilly}}]{biancalana2007dynamics}%
  \BibitemOpen
  \bibfield  {author} {\bibinfo {author} {\bibfnamefont {F.}~\bibnamefont {Biancalana}}, \bibinfo {author} {\bibfnamefont {A.}~\bibnamefont {Amann}}, \bibinfo {author} {\bibfnamefont {A.~V.}\ \bibnamefont {Uskov}},\ and\ \bibinfo {author} {\bibfnamefont {E.~P.}\ \bibnamefont {O'Reilly}},\ }\bibfield  {title} {\bibinfo {title} {Dynamics of light propagation in spatiotemporal dielectric structures},\ }\href {https://doi.org/10.1103/PhysRevE.75.046607} {\bibfield  {journal} {\bibinfo  {journal} {Phys. Rev. E}\ }\textbf {\bibinfo {volume} {75}},\ \bibinfo {pages} {046607} (\bibinfo {year} {2007})}\BibitemShut {NoStop}%
\bibitem [{\citenamefont {Cong}\ and\ \citenamefont {Singh}(2020)}]{cong2020spatiotemporal}%
  \BibitemOpen
  \bibfield  {author} {\bibinfo {author} {\bibfnamefont {L.}~\bibnamefont {Cong}}\ and\ \bibinfo {author} {\bibfnamefont {R.}~\bibnamefont {Singh}},\ }\bibfield  {title} {\bibinfo {title} {Spatiotemporal dielectric metasurfaces for unidirectional propagation and reconfigurable steering of terahertz beams},\ }\href {https://doi.org/https://doi.org/10.1002/adma.202001418} {\bibfield  {journal} {\bibinfo  {journal} {Adv. Mater.}\ }\textbf {\bibinfo {volume} {32}},\ \bibinfo {pages} {2001418} (\bibinfo {year} {2020})}\BibitemShut {NoStop}%
\bibitem [{\citenamefont {Sharabi}\ \emph {et~al.}(2022)\citenamefont {Sharabi}, \citenamefont {Dikopoltsev}, \citenamefont {Lustig}, \citenamefont {Lumer},\ and\ \citenamefont {Segev}}]{sharabi2022spatiotemporal}%
  \BibitemOpen
  \bibfield  {author} {\bibinfo {author} {\bibfnamefont {Y.}~\bibnamefont {Sharabi}}, \bibinfo {author} {\bibfnamefont {A.}~\bibnamefont {Dikopoltsev}}, \bibinfo {author} {\bibfnamefont {E.}~\bibnamefont {Lustig}}, \bibinfo {author} {\bibfnamefont {Y.}~\bibnamefont {Lumer}},\ and\ \bibinfo {author} {\bibfnamefont {M.}~\bibnamefont {Segev}},\ }\bibfield  {title} {\bibinfo {title} {Spatiotemporal photonic crystals},\ }\href {https://doi.org/10.1364/OPTICA.455672} {\bibfield  {journal} {\bibinfo  {journal} {Optica}\ }\textbf {\bibinfo {volume} {9}},\ \bibinfo {pages} {585} (\bibinfo {year} {2022})}\BibitemShut {NoStop}%
\bibitem [{\citenamefont {Park}\ and\ \citenamefont {Min}(2021)}]{park2021spatiotemporal}%
  \BibitemOpen
  \bibfield  {author} {\bibinfo {author} {\bibfnamefont {J.}~\bibnamefont {Park}}\ and\ \bibinfo {author} {\bibfnamefont {B.}~\bibnamefont {Min}},\ }\bibfield  {title} {\bibinfo {title} {Spatiotemporal plane wave expansion method for arbitrary space--time periodic photonic media},\ }\href {https://doi.org/10.1364/OL.411622} {\bibfield  {journal} {\bibinfo  {journal} {Opt. Lett.}\ }\textbf {\bibinfo {volume} {46}},\ \bibinfo {pages} {484} (\bibinfo {year} {2021})}\BibitemShut {NoStop}%
\bibitem [{\citenamefont {Ding}\ and\ \citenamefont {Ding}(2024)}]{ding2024non}%
  \BibitemOpen
  \bibfield  {author} {\bibinfo {author} {\bibfnamefont {H.}~\bibnamefont {Ding}}\ and\ \bibinfo {author} {\bibfnamefont {K.}~\bibnamefont {Ding}},\ }\bibfield  {title} {\bibinfo {title} {{Non-Bloch} theory for spatiotemporal photonic crystals assisted by continuum effective medium},\ }\href {https://doi.org/10.1103/PhysRevResearch.6.033167} {\bibfield  {journal} {\bibinfo  {journal} {Phys. Rev. Res.}\ }\textbf {\bibinfo {volume} {6}},\ \bibinfo {pages} {033167} (\bibinfo {year} {2024})}\BibitemShut {NoStop}%
\bibitem [{\citenamefont {Galiffi}\ \emph {et~al.}(2022)\citenamefont {Galiffi}, \citenamefont {Tirole}, \citenamefont {Yin}, \citenamefont {Li}, \citenamefont {Vezzoli}, \citenamefont {Huidobro}, \citenamefont {Silveirinha}, \citenamefont {Sapienza}, \citenamefont {Al{\`u}},\ and\ \citenamefont {Pendry}}]{galiffi2022photonics}%
  \BibitemOpen
  \bibfield  {author} {\bibinfo {author} {\bibfnamefont {E.}~\bibnamefont {Galiffi}}, \bibinfo {author} {\bibfnamefont {R.}~\bibnamefont {Tirole}}, \bibinfo {author} {\bibfnamefont {S.}~\bibnamefont {Yin}}, \bibinfo {author} {\bibfnamefont {H.}~\bibnamefont {Li}}, \bibinfo {author} {\bibfnamefont {S.}~\bibnamefont {Vezzoli}}, \bibinfo {author} {\bibfnamefont {P.~A.}\ \bibnamefont {Huidobro}}, \bibinfo {author} {\bibfnamefont {M.~G.}\ \bibnamefont {Silveirinha}}, \bibinfo {author} {\bibfnamefont {R.}~\bibnamefont {Sapienza}}, \bibinfo {author} {\bibfnamefont {A.}~\bibnamefont {Al{\`u}}},\ and\ \bibinfo {author} {\bibfnamefont {J.~B.}\ \bibnamefont {Pendry}},\ }\bibfield  {title} {\bibinfo {title} {Photonics of time-varying media},\ }\href {https://doi.org/10.1117/1.AP.4.1.014002} {\bibfield  {journal} {\bibinfo  {journal} {Adv. Photonics}\ }\textbf {\bibinfo {volume} {4}},\ \bibinfo {pages} {014002} (\bibinfo {year} {2022})}\BibitemShut {NoStop}%
\bibitem [{\citenamefont {Sounas}\ and\ \citenamefont {Al{\`u}}(2017)}]{sounas2017non}%
  \BibitemOpen
  \bibfield  {author} {\bibinfo {author} {\bibfnamefont {D.~L.}\ \bibnamefont {Sounas}}\ and\ \bibinfo {author} {\bibfnamefont {A.}~\bibnamefont {Al{\`u}}},\ }\bibfield  {title} {\bibinfo {title} {Non-reciprocal photonics based on time modulation},\ }\href {https://doi.org/10.1038/s41566-017-0051-x} {\bibfield  {journal} {\bibinfo  {journal} {Nat. Photonics}\ }\textbf {\bibinfo {volume} {11}},\ \bibinfo {pages} {774} (\bibinfo {year} {2017})}\BibitemShut {NoStop}%
\bibitem [{\citenamefont {Kazemi}\ \emph {et~al.}(2019)\citenamefont {Kazemi}, \citenamefont {Nada}, \citenamefont {Mealy}, \citenamefont {Abdelshafy},\ and\ \citenamefont {Capolino}}]{kazemi2019exceptional}%
  \BibitemOpen
  \bibfield  {author} {\bibinfo {author} {\bibfnamefont {H.}~\bibnamefont {Kazemi}}, \bibinfo {author} {\bibfnamefont {M.~Y.}\ \bibnamefont {Nada}}, \bibinfo {author} {\bibfnamefont {T.}~\bibnamefont {Mealy}}, \bibinfo {author} {\bibfnamefont {A.~F.}\ \bibnamefont {Abdelshafy}},\ and\ \bibinfo {author} {\bibfnamefont {F.}~\bibnamefont {Capolino}},\ }\bibfield  {title} {\bibinfo {title} {Exceptional points of degeneracy induced by linear time-periodic variation},\ }\href {https://doi.org/10.1103/PhysRevApplied.11.014007} {\bibfield  {journal} {\bibinfo  {journal} {Phys. Rev. Appl.}\ }\textbf {\bibinfo {volume} {11}},\ \bibinfo {pages} {014007} (\bibinfo {year} {2019})}\BibitemShut {NoStop}%
\bibitem [{\citenamefont {Lustig}\ \emph {et~al.}(2018)\citenamefont {Lustig}, \citenamefont {Sharabi},\ and\ \citenamefont {Segev}}]{lustig2018topological}%
  \BibitemOpen
  \bibfield  {author} {\bibinfo {author} {\bibfnamefont {E.}~\bibnamefont {Lustig}}, \bibinfo {author} {\bibfnamefont {Y.}~\bibnamefont {Sharabi}},\ and\ \bibinfo {author} {\bibfnamefont {M.}~\bibnamefont {Segev}},\ }\bibfield  {title} {\bibinfo {title} {Topological aspects of photonic time crystals},\ }\href {https://doi.org/10.1364/OPTICA.5.001390} {\bibfield  {journal} {\bibinfo  {journal} {Optica}\ }\textbf {\bibinfo {volume} {5}},\ \bibinfo {pages} {1390} (\bibinfo {year} {2018})}\BibitemShut {NoStop}%
\bibitem [{\citenamefont {Lu}\ \emph {et~al.}(2021)\citenamefont {Lu}, \citenamefont {He}, \citenamefont {Addison}, \citenamefont {Mele},\ and\ \citenamefont {Zhen}}]{lu2021floquet}%
  \BibitemOpen
  \bibfield  {author} {\bibinfo {author} {\bibfnamefont {J.}~\bibnamefont {Lu}}, \bibinfo {author} {\bibfnamefont {L.}~\bibnamefont {He}}, \bibinfo {author} {\bibfnamefont {Z.}~\bibnamefont {Addison}}, \bibinfo {author} {\bibfnamefont {E.~J.}\ \bibnamefont {Mele}},\ and\ \bibinfo {author} {\bibfnamefont {B.}~\bibnamefont {Zhen}},\ }\bibfield  {title} {\bibinfo {title} {Floquet topological phases in one-dimensional nonlinear photonic crystals},\ }\href {https://doi.org/10.1103/PhysRevLett.126.113901} {\bibfield  {journal} {\bibinfo  {journal} {Phys. Rev. Lett.}\ }\textbf {\bibinfo {volume} {126}},\ \bibinfo {pages} {113901} (\bibinfo {year} {2021})}\BibitemShut {NoStop}%
\bibitem [{\citenamefont {Xiao}\ \emph {et~al.}(2014)\citenamefont {Xiao}, \citenamefont {Zhang},\ and\ \citenamefont {Chan}}]{xiao2014surface}%
  \BibitemOpen
  \bibfield  {author} {\bibinfo {author} {\bibfnamefont {M.}~\bibnamefont {Xiao}}, \bibinfo {author} {\bibfnamefont {Z.}~\bibnamefont {Zhang}},\ and\ \bibinfo {author} {\bibfnamefont {C.~T.}\ \bibnamefont {Chan}},\ }\bibfield  {title} {\bibinfo {title} {Surface impedance and bulk band geometric phases in one-dimensional systems},\ }\href {https://doi.org/10.1103/PhysRevX.4.021017} {\bibfield  {journal} {\bibinfo  {journal} {Phys. Rev. X}\ }\textbf {\bibinfo {volume} {4}},\ \bibinfo {pages} {021017} (\bibinfo {year} {2014})}\BibitemShut {NoStop}%
\bibitem [{\citenamefont {Ding}\ \emph {et~al.}(2015)\citenamefont {Ding}, \citenamefont {Zhang},\ and\ \citenamefont {Chan}}]{ding2015coalescence}%
  \BibitemOpen
  \bibfield  {author} {\bibinfo {author} {\bibfnamefont {K.}~\bibnamefont {Ding}}, \bibinfo {author} {\bibfnamefont {Z.}~\bibnamefont {Zhang}},\ and\ \bibinfo {author} {\bibfnamefont {C.~T.}\ \bibnamefont {Chan}},\ }\bibfield  {title} {\bibinfo {title} {Coalescence of exceptional points and phase diagrams for one-dimensional $\mathcal{P}\mathcal{T}$-symmetric photonic crystals},\ }\href {https://doi.org/10.1103/PhysRevB.92.235310} {\bibfield  {journal} {\bibinfo  {journal} {Phys. Rev. B}\ }\textbf {\bibinfo {volume} {92}},\ \bibinfo {pages} {235310} (\bibinfo {year} {2015})}\BibitemShut {NoStop}%
\bibitem [{\citenamefont {Yao}\ and\ \citenamefont {Wang}(2018)}]{yao2018edge}%
  \BibitemOpen
  \bibfield  {author} {\bibinfo {author} {\bibfnamefont {S.}~\bibnamefont {Yao}}\ and\ \bibinfo {author} {\bibfnamefont {Z.}~\bibnamefont {Wang}},\ }\bibfield  {title} {\bibinfo {title} {Edge states and topological invariants of {non-Hermitian} systems},\ }\href {https://doi.org/10.1103/PhysRevLett.121.086803} {\bibfield  {journal} {\bibinfo  {journal} {Phys. Rev. Lett.}\ }\textbf {\bibinfo {volume} {121}},\ \bibinfo {pages} {086803} (\bibinfo {year} {2018})}\BibitemShut {NoStop}%
\bibitem [{\citenamefont {Song}\ \emph {et~al.}(2019)\citenamefont {Song}, \citenamefont {Yao},\ and\ \citenamefont {Wang}}]{song2019non}%
  \BibitemOpen
  \bibfield  {author} {\bibinfo {author} {\bibfnamefont {F.}~\bibnamefont {Song}}, \bibinfo {author} {\bibfnamefont {S.}~\bibnamefont {Yao}},\ and\ \bibinfo {author} {\bibfnamefont {Z.}~\bibnamefont {Wang}},\ }\bibfield  {title} {\bibinfo {title} {{Non-Hermitian} topological invariants in real space},\ }\href {https://doi.org/10.1103/PhysRevLett.123.246801} {\bibfield  {journal} {\bibinfo  {journal} {Phys. Rev. Lett.}\ }\textbf {\bibinfo {volume} {123}},\ \bibinfo {pages} {246801} (\bibinfo {year} {2019})}\BibitemShut {NoStop}%
\bibitem [{\citenamefont {Yokomizo}\ and\ \citenamefont {Murakami}(2019)}]{yokomizo2019non}%
  \BibitemOpen
  \bibfield  {author} {\bibinfo {author} {\bibfnamefont {K.}~\bibnamefont {Yokomizo}}\ and\ \bibinfo {author} {\bibfnamefont {S.}~\bibnamefont {Murakami}},\ }\bibfield  {title} {\bibinfo {title} {{Non-Bloch} band theory of {non-Hermitian} systems},\ }\href {https://doi.org/10.1103/PhysRevLett.123.066404} {\bibfield  {journal} {\bibinfo  {journal} {Phys. Rev. Lett.}\ }\textbf {\bibinfo {volume} {123}},\ \bibinfo {pages} {066404} (\bibinfo {year} {2019})}\BibitemShut {NoStop}%
\bibitem [{\citenamefont {Yokomizo}\ \emph {et~al.}(2022)\citenamefont {Yokomizo}, \citenamefont {Yoda},\ and\ \citenamefont {Murakami}}]{yokomizo2022non}%
  \BibitemOpen
  \bibfield  {author} {\bibinfo {author} {\bibfnamefont {K.}~\bibnamefont {Yokomizo}}, \bibinfo {author} {\bibfnamefont {T.}~\bibnamefont {Yoda}},\ and\ \bibinfo {author} {\bibfnamefont {S.}~\bibnamefont {Murakami}},\ }\bibfield  {title} {\bibinfo {title} {{Non-Hermitian} waves in a continuous periodic model and application to photonic crystals},\ }\href {https://doi.org/10.1103/PhysRevResearch.4.023089} {\bibfield  {journal} {\bibinfo  {journal} {Phys. Rev. Res.}\ }\textbf {\bibinfo {volume} {4}},\ \bibinfo {pages} {023089} (\bibinfo {year} {2022})}\BibitemShut {NoStop}%
\bibitem [{\citenamefont {Yokomizo}\ \emph {et~al.}(2024)\citenamefont {Yokomizo}, \citenamefont {Yoda},\ and\ \citenamefont {Ashida}}]{yokomizo2024non}%
  \BibitemOpen
  \bibfield  {author} {\bibinfo {author} {\bibfnamefont {K.}~\bibnamefont {Yokomizo}}, \bibinfo {author} {\bibfnamefont {T.}~\bibnamefont {Yoda}},\ and\ \bibinfo {author} {\bibfnamefont {Y.}~\bibnamefont {Ashida}},\ }\bibfield  {title} {\bibinfo {title} {{Non-Bloch} band theory of generalized eigenvalue problems},\ }\href {https://doi.org/10.1103/PhysRevB.109.115115} {\bibfield  {journal} {\bibinfo  {journal} {Phys. Rev. B}\ }\textbf {\bibinfo {volume} {109}},\ \bibinfo {pages} {115115} (\bibinfo {year} {2024})}\BibitemShut {NoStop}%
\bibitem [{\citenamefont {Huidobro}\ \emph {et~al.}(2021)\citenamefont {Huidobro}, \citenamefont {Silveirinha}, \citenamefont {Galiffi},\ and\ \citenamefont {Pendry}}]{huidobro2021homogenization}%
  \BibitemOpen
  \bibfield  {author} {\bibinfo {author} {\bibfnamefont {P.~A.}\ \bibnamefont {Huidobro}}, \bibinfo {author} {\bibfnamefont {M.~G.}\ \bibnamefont {Silveirinha}}, \bibinfo {author} {\bibfnamefont {E.}~\bibnamefont {Galiffi}},\ and\ \bibinfo {author} {\bibfnamefont {J.}~\bibnamefont {Pendry}},\ }\bibfield  {title} {\bibinfo {title} {Homogenization theory of space-time metamaterials},\ }\href {https://doi.org/10.1103/PhysRevApplied.16.014044} {\bibfield  {journal} {\bibinfo  {journal} {Phys. Rev. Appl.}\ }\textbf {\bibinfo {volume} {16}},\ \bibinfo {pages} {014044} (\bibinfo {year} {2021})}\BibitemShut {NoStop}%
\bibitem [{sup()}]{supplementary}%
  \BibitemOpen
  \href@noop {} {}\bibinfo {note} {Supplemental Material is available at [url].}\BibitemShut {Stop}%
\bibitem [{\citenamefont {Blanco~de Paz}\ \emph {et~al.}(2020)\citenamefont {Blanco~de Paz}, \citenamefont {Devescovi}, \citenamefont {Giedke}, \citenamefont {Saenz}, \citenamefont {Vergniory}, \citenamefont {Bradlyn}, \citenamefont {Bercioux},\ and\ \citenamefont {Garc{\'i}a-Etxarri}}]{blanco2020tutorial}%
  \BibitemOpen
  \bibfield  {author} {\bibinfo {author} {\bibfnamefont {M.}~\bibnamefont {Blanco~de Paz}}, \bibinfo {author} {\bibfnamefont {C.}~\bibnamefont {Devescovi}}, \bibinfo {author} {\bibfnamefont {G.}~\bibnamefont {Giedke}}, \bibinfo {author} {\bibfnamefont {J.~J.}\ \bibnamefont {Saenz}}, \bibinfo {author} {\bibfnamefont {M.~G.}\ \bibnamefont {Vergniory}}, \bibinfo {author} {\bibfnamefont {B.}~\bibnamefont {Bradlyn}}, \bibinfo {author} {\bibfnamefont {D.}~\bibnamefont {Bercioux}},\ and\ \bibinfo {author} {\bibfnamefont {A.}~\bibnamefont {Garc{\'i}a-Etxarri}},\ }\bibfield  {title} {\bibinfo {title} {Tutorial: computing topological invariants in {2D} photonic crystals},\ }\href {https://doi.org/10.1002/qute.201900117} {\bibfield  {journal} {\bibinfo  {journal} {Adv. Quantum Technol.}\ }\textbf {\bibinfo {volume} {3}},\ \bibinfo {pages} {1900117} (\bibinfo {year} {2020})}\BibitemShut {NoStop}%
\bibitem [{\citenamefont {Stegmaier}\ \emph {et~al.}(2021)\citenamefont {Stegmaier}, \citenamefont {Imhof}, \citenamefont {Helbig}, \citenamefont {Hofmann}, \citenamefont {Lee}, \citenamefont {Kremer}, \citenamefont {Fritzsche}, \citenamefont {Feichtner}, \citenamefont {Klembt}, \citenamefont {H{\"o}fling} \emph {et~al.}}]{stegmaier2021topological}%
  \BibitemOpen
  \bibfield  {author} {\bibinfo {author} {\bibfnamefont {A.}~\bibnamefont {Stegmaier}}, \bibinfo {author} {\bibfnamefont {S.}~\bibnamefont {Imhof}}, \bibinfo {author} {\bibfnamefont {T.}~\bibnamefont {Helbig}}, \bibinfo {author} {\bibfnamefont {T.}~\bibnamefont {Hofmann}}, \bibinfo {author} {\bibfnamefont {C.~H.}\ \bibnamefont {Lee}}, \bibinfo {author} {\bibfnamefont {M.}~\bibnamefont {Kremer}}, \bibinfo {author} {\bibfnamefont {A.}~\bibnamefont {Fritzsche}}, \bibinfo {author} {\bibfnamefont {T.}~\bibnamefont {Feichtner}}, \bibinfo {author} {\bibfnamefont {S.}~\bibnamefont {Klembt}}, \bibinfo {author} {\bibfnamefont {S.}~\bibnamefont {H{\"o}fling}}, \emph {et~al.},\ }\bibfield  {title} {\bibinfo {title} {Topological defect engineering and $\mathcal{P}\mathcal{T}$ symmetry in {non-Hermitian} electrical circuits},\ }\href {https://doi.org/10.1103/PhysRevLett.126.215302} {\bibfield  {journal} {\bibinfo  {journal} {Phys. Rev. Lett.}\ }\textbf {\bibinfo {volume} {126}},\ \bibinfo {pages} {215302} (\bibinfo {year}
  {2021})}\BibitemShut {NoStop}%
\bibitem [{\citenamefont {Yang}\ \emph {et~al.}(2024)\citenamefont {Yang}, \citenamefont {Yang}, \citenamefont {Ma}, \citenamefont {Li}, \citenamefont {Zhang},\ and\ \citenamefont {Chan}}]{yang2024non}%
  \BibitemOpen
  \bibfield  {author} {\bibinfo {author} {\bibfnamefont {Y.}~\bibnamefont {Yang}}, \bibinfo {author} {\bibfnamefont {B.}~\bibnamefont {Yang}}, \bibinfo {author} {\bibfnamefont {G.}~\bibnamefont {Ma}}, \bibinfo {author} {\bibfnamefont {J.}~\bibnamefont {Li}}, \bibinfo {author} {\bibfnamefont {S.}~\bibnamefont {Zhang}},\ and\ \bibinfo {author} {\bibfnamefont {C.~T.}\ \bibnamefont {Chan}},\ }\bibfield  {title} {\bibinfo {title} {{Non-Abelian} physics in light and sound},\ }\href {https://doi.org/10.1126/science.adf9621} {\bibfield  {journal} {\bibinfo  {journal} {Science}\ }\textbf {\bibinfo {volume} {383}},\ \bibinfo {pages} {eadf9621} (\bibinfo {year} {2024})}\BibitemShut {NoStop}%
\bibitem [{\citenamefont {Zak}(1989)}]{zak1989berry}%
  \BibitemOpen
  \bibfield  {author} {\bibinfo {author} {\bibfnamefont {J.}~\bibnamefont {Zak}},\ }\bibfield  {title} {\bibinfo {title} {Berry's phase for energy bands in solids},\ }\href {https://doi.org/10.1103/PhysRevLett.62.2747} {\bibfield  {journal} {\bibinfo  {journal} {Phys. Rev. Lett.}\ }\textbf {\bibinfo {volume} {62}},\ \bibinfo {pages} {2747} (\bibinfo {year} {1989})}\BibitemShut {NoStop}%
\bibitem [{\citenamefont {Xiao}\ \emph {et~al.}(2015)\citenamefont {Xiao}, \citenamefont {Ma}, \citenamefont {Yang}, \citenamefont {Sheng}, \citenamefont {Zhang},\ and\ \citenamefont {Chan}}]{xiao2015geometric}%
  \BibitemOpen
  \bibfield  {author} {\bibinfo {author} {\bibfnamefont {M.}~\bibnamefont {Xiao}}, \bibinfo {author} {\bibfnamefont {G.}~\bibnamefont {Ma}}, \bibinfo {author} {\bibfnamefont {Z.}~\bibnamefont {Yang}}, \bibinfo {author} {\bibfnamefont {P.}~\bibnamefont {Sheng}}, \bibinfo {author} {\bibfnamefont {Z.~Q.}\ \bibnamefont {Zhang}},\ and\ \bibinfo {author} {\bibfnamefont {C.~T.}\ \bibnamefont {Chan}},\ }\bibfield  {title} {\bibinfo {title} {Geometric phase and band inversion in periodic acoustic systems},\ }\href {https://doi.org/10.1038/nphys3228} {\bibfield  {journal} {\bibinfo  {journal} {Nat. Phys.}\ }\textbf {\bibinfo {volume} {11}},\ \bibinfo {pages} {240} (\bibinfo {year} {2015})}\BibitemShut {NoStop}%
\end{thebibliography}%

\end{document}